\documentclass[aps,prl,twocolumn,superscriptaddress,showpacs]{revtex4}

\usepackage{graphics}
\usepackage{txfonts}

\newcommand{\cps}{CePt$_3$Si}
\newcommand{\lps}{LaPt$_3$Si}
\begin{document}

\title{Magnetic penetration depth and gap symmetry
   of the noncentrosymmetric superconductors CePt$_3$Si and
   LaPt$_3$Si}

\author{R. L. Ribeiro}
\affiliation{Centro de F\'{\i}sica, Instituto Venezolano de
Investigaciones Cient\'{\i}ficas, Apartado 20632, Caracas
1020-A, Venezuela}

\author{I. Bonalde}
\affiliation{Centro de F\'{\i}sica, Instituto Venezolano de
Investigaciones Cient\'{\i}ficas, Apartado 20632, Caracas
1020-A, Venezuela}. \affiliation{Center for Quantum Science and
Technology under Extreme Conditions and Graduate School of
Science, Osaka University, Toyonaka, Osaka 560-0043, Japan}

\author{Y. Haga}
\affiliation{Advanced Science Research Center, Japan Atomic Energy
Research Institute, Tokai, Ibaragi 319-1195, Japan}

\author{R. Settai}
\affiliation{Graduate School of Science, Osaka University,
Toyonaka, Osaka 560-0043, Japan}

\author{Y. \={O}nuki}
\affiliation{Advanced Science Research Center, Japan Atomic
Energy Research Institute, Tokai, Ibaragi 319-1195,
Japan}.\affiliation{Graduate School of Science, Osaka
University, Toyonaka, Osaka 560-0043, Japan}

\begin{abstract}

\end{abstract}


\maketitle

The role of broken parity in the unconventional responses of
superconductors without inversion symmetry has been difficult
to pinpoint. The absence of inversion symmetry in a crystal
structure causes the appearance of an antisymmetric spin-orbit
coupling (ASOC) that affects the electronic properties. Some of
these superconductors, like \cps\ and the series CeTX$_3$
(T=Rh, Ir, Co; X=Si, Ge), are also antiferromagnetic
heavy-fermions. A few models based on the lack of parity
\cite{samokhin,sergienko,frigeri} and the effect of
antiferromagnetic order \cite{fujimoto2,yanase1} have been
introduced mainly to describe the unconventional behaviors of
\cps.\ Among such behaviors are line nodes in the gap
\cite{mine7,izawa,takeuchi1}, an upper-critical field larger
than the paramagnetic limiting field \cite{bauer,takeuchi} and
a constant spin susceptibility across the transition
\cite{yogi2}. The models are supposed to explain all
superconductors without inversion symmetry, but most of the
nonmagnetic noncentrosymmetric superconductors with a strong
ASOC display conventional $s$-wave superconductivity. Thus,
there is an uncertainty on what is really causing both types of
behaviors. This doubt calls for further studies addressing the
origin of unconventional responses in superconductors without
inversion symmetry.

Here, we aim to get further insight into the importance of the
lack of parity in the unusual responses of CePt$_3$Si by
studying the isostructural LaPt$_3$Si ($T_c=0.64$ K) without
electron correlations. LaPt$_3$Si and CePt$_3$Si have similar
ASOC strengths \cite{samokhin,mineev1}. Thus, LaPt$_3$Si
constitutes a special system to test the role of both electron
correlations and broken parity in CePt$_3$Si. The
superconducting phase of LaPt$_3$Si has been hardly studied.
Specific-heat data suggest that LaPt$_3$Si is a weak-coupling
$s$-wave superconductor \cite{takeuchi1}. Experimental results
in other superconducting properties are then required to
confirm this. We report here on high-resolution magnetic
penetration depth $\lambda(T)$ measurements of a high-quality
single crystal of LaPt$_3$Si down to 60 mK ($\sim 0.09T_c$). We
found a broad superconducting transition and evidence for
conventional $s$-wave superconductivity.

The single crystal of LaPt$_3$Si used in our experiment was
grown by the Bridgman method \cite{takeuchi1} and has
dimensions around $0.43\times0.48\times0.28$ mm$^3$. The
observation of de Haas-van Alphen oscillations and the
mean-free path values as large as 2400 $\AA$ in single crystals
of the same batch \cite{hashimoto1} are strong indications of
the high-quality of our crystal. Penetration depth measurements
were performed utilizing a 13 MHz tunnel diode oscillator
\cite{mine7}. The magnitude of the ac probing field was
estimated to be 3 mOe, and the dc field at the sample was
reduced to around 1 mOe. The deviation of the penetration depth
from the lowest measured temperature,
$\Delta\lambda(T)=\lambda(T)-\lambda$(0.06 K), was obtained up
to $T \sim 0.99T_c$ from the change in the measured resonance
frequency $\Delta f(T) = G \Delta \lambda(T)$. No difference is
observed by using $\Delta f(T) = G \Delta \chi(T)$, with the
full sample susceptibility
$\chi=[(2\lambda/a)\textrm{tanh}(a/2\lambda)-1]$ for a slab
\cite{mine7}. Here $G$ is a constant factor that depends on the
sample and coil geometries and that includes the demagnetizing
factor of the sample, and $a$ is the relevant dimension.

%
\begin{figure}
\begin{center}\scalebox{0.8}{\includegraphics{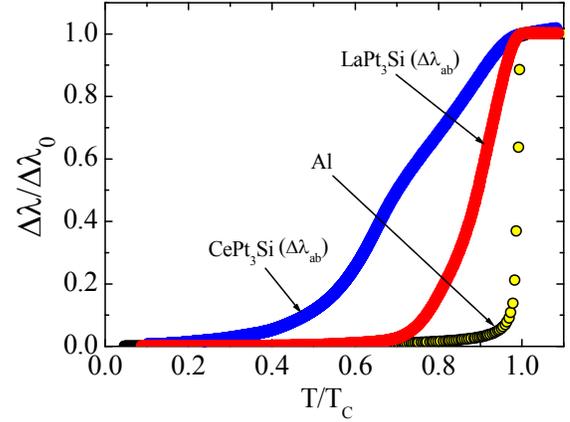}}
\caption{(Color online) Normalized
$\Delta\lambda_{ab}(T)/\Delta\lambda_0$ of LaPt$_3$Si and
single crystal B2 of CePt$_3$Si reported in ref.~14. For
comparison the normalized variation of the penetration depth of
aluminum is also displayed.} \label{lambdatotal}
\end{center}
\end{figure}
%
%

Figure~\ref{lambdatotal} shows the normalized variation
$\Delta\lambda_{ab}(T)/\Delta\lambda_0$ as a function of
temperature in LaPt$_3$Si and single crystal B2 of CePt$_3$Si
reported in ref.~14. Here $\Delta\lambda_0$ is the total
penetration depth shift of the samples. For comparison, the
figure also depicts the normalized variation in the penetration
depth of a 99.9995$\%$ Al polycrystalline sample. In \lps\ the
out-of-plane penetration depth $\lambda_{c}(T)$ is similar to
$\lambda_{ab}(T)$, which implies isotropic superconductivity as
in CePt$_3$Si. A noticeable feature of the figure is that the
superconducting transition of LaPt$_3$Si is not sharp
(transition width around 0.15 K), as expected for a very clean
high-quality single crystal. The transition is also quite broad
in CePt$_3$Si \cite{mine13}. Interestingly, a linear
temperature response of the penetration depth of CePt$_3$Si and
LaPt$_3$Si is clearly seen in the region just below $T_c$. The
wide transition would lead to a strong suppression of the
superfluid density near $T_c$ in LaPt$_3$Si, as observed in
CePt$_3$Si \cite{mine7,mine9}. A broad superconducting
transition in LaPt$_3$Si was also observed, even tough not
discussed, in specific-heat \cite{takeuchi1} and resistivity
\cite{takeuchipc} measurements. In LaPt$_3$Si and CePt$_3$Si
the wide transitions should not be associated with impurities
or defects. The fact that the penetration depth data of
LaPt$_3$Si do not show any kink or second drop suggests that
there are no magnetic or second superconducting transitions
below $T_c$. The broad transition in LaPt$_3$Si can be
evidently attributed to neither magnetic nor heavy-fermion
effects. It is unlikely that the wide transitions are a
signature of broken parity, since other noncentrosymmetric
superconductors show very sharp transitions. The
superconducting transition in CePt$_3$Si has been found to
sharpen around the pressure at which the antiferromagnetic
phase disappears \cite{aoki2}.

%
%
\begin{figure}
\begin{center}
\scalebox{0.8}{\includegraphics{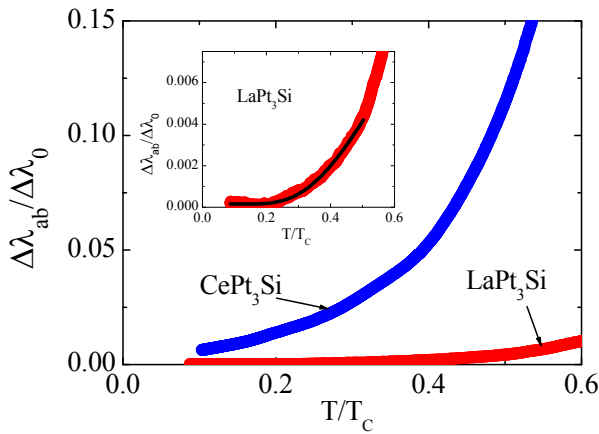}} \caption{(Color
online)  Low-temperature region of the variation in the
penetration depth of LaPt$_3$Si and CePt$_3$Si. The data of
LaPt$_3$Si follow a BCS $s$-wave behavior (see inset), whereas
those of CePt$_3$Si response linearly as $T \rightarrow 0$. The
solid line in the inset is a fit to the BCS model for an
isotropic pairing symmetry.}\label{lambdalowT}
\end{center}
\end{figure}
%
%

The main body of Fig.~\ref{lambdalowT} displays the
low-temperature region of the penetration depth data of
LaPt$_3$Si and CePt$_3$Si shown in Fig.~\ref{lambdatotal}.
Whereas in CePt$_3$Si the penetration depth changes linearly
with temperature as $T\rightarrow 0$, indicating line nodes in
the gap \cite{mine7,mine13}, in LaPt$_3$Si it flattens out
below about 0.2$T_c$ (see inset to Fig.~\ref{lambdalowT}), as
theoretically expected for a superconductor with an isotropic
energy gap. At temperatures $T<0.5T_c$ the data of LaPt$_3$Si
are fitted very well to the BCS model
\begin{equation}
\label{swavebcs} \Delta\lambda(T) \propto \sqrt{\frac{\pi
\Delta_0}{2k_BT}} \exp(-\Delta_0/k_BT) \, ,
\end{equation}
with $\Delta_0=1.73k_BT_c$. This value is quite similar to that
of the weak-coupling BCS model $1.76k_BT_c$. Overall the
behavior of the penetration depth of LaPt$_3$Si is in agreement
with that observed in specific-heat
measurements~\cite{takeuchi1}, although in the latter case a
lower value $\Delta_0=1.35k_BT_c$ was obtained.

The isostructural LaPt$_3$Si and CePt$_3$Si, apart from having
similar ASOC strengths, are thought to have similar Fermi
surfaces and contributions of the bands to the density of
states \cite{hashimoto1}. If parity mixing alone determines the
behaviors of the superconducting properties in these compounds,
one would expect such behaviors to be similar. However,
LaPt$_3$Si presents conventional $s$-wave responses, while
CePt$_3$Si shows unconventional ones. Similar situation appears
to take place in isostructural LaIrSi$_3$ and CeIrSi$_3$, for
which NMR measurements suggest an isotropic gap and a gap with
line nodes, respectively \cite{mukuda1}. Thus, it seems that in
CePt$_3$Si and CeIrSi$_3$ parity mixing does not lead -at least
solely- to line nodes. For the case of CePt$_3$Si (applicable
to CeIrSi$_3$), it has been pointed out that line nodes are
accidentally generated when the antiferromagnetic order is
taken into account and the $p$-wave component of the parity
mixing is dominant \cite{fujimoto2,yanase1}. Morever, it is
thought that the antiferromagnetic order favors the $p$-wave
component and that in the absence of such ordering the $s$-wave
component dominates. This will be consistent with the present
experimental results. Within this theoretical scenario the
nonmagnetic Li$_2$Pt$_3$B with line nodes remains a puzzle,
being the only nonmagnetic superconductor with a strong ASOC
that does not display a BCS $s$-wave behavior. In a different
scheme line nodes in noncentrosymmetric compounds can still be
imposed by symmetry \cite{samokhin,sergienko}, as it occurs in
some unconventional superconductors with inversion symmetry.

In summary, from measurements of the magnetic penetration depth
we found that the superconducting transition of LaPt$_3$Si is
broad, as is in CePt$_3$Si. We also observed a conventional
$s$-wave behavior in LaPt$_3$Si, as opposed to the
unconventional response obtained in CePt$_3$Si. Since these
compounds are isostructural and have the same ASOC strength,
the present result would imply that the parity mixing alone
does not lead to unconventional behaviors in \cps\ and that the
antiferromagnetic order may need to be taken into account.

\section*{Acknowledgment}

We highly appreciate discussions with T. Takeuchi and valuable
comments from M. Sigrist, D. F. Agterberg, K. Samokhin, and S.
Fujimoto.


\end{document}